# Free Standing Layers of Silicene like Sheets

*Shobha Shukla, Raghvendra Pratap Chaudhary and Sumit Saxena[*]*
*Nanostructures engineering and modeling laboratory, Department of Metallurgical Engineering and Materials Science, Indian Institute of Technology Bombay, Mumbai, India 400076*

[*] Corresponding author sumit.saxena@iitb.ac.in

**Abstract**

Silicene is predicted to possess exotic electronic properties and a forerunner amongst all 2D materials for the development of exotic devices using present silicon technology. Here we report the synthesis of free standing layered 2D hexagonal silicene like sheets. Atomic scale structural characterizations have been performed using HRTEM. Samples have been characterized using HRXRD, SAED, UV-Vis absorption and Raman measurements, which suggests that the samples are silicene like sheets and not exfoliated (111) silicon sheets.

Unique symmetry of π orbital network in hexagonal lattice enables group IV layered materials to exhibit exotic electronic properties. The discovery of graphene in 2004 followed by its extraordinary electronic, mechanical and optical properties has created a runaway effect in exploration of such layered 2D materials[1]. Silicon and Carbon enjoy similar valence electronic configuration by the virtue of both lying in the group IV of the periodic table. In this perspective silicene, a mono atomic thin layer of hexagonal silicon has been projected as graphene's cousin in terms of its exotic properties. A lot of experimental activity catalyzed by fundamental theoretical investigations has resulted in realization of epitaxial growth of silicene sheets. Growth studies suggest that the surface morphology of silicene is substrate dependent. It is observed that silicene changes structural forms on varying the substrate temperatures on Ag (111) surface[2], forms corrugated surface on Ir(111)[3] and $ZrB_2(0001)$ surface[4]. This suggest that honeycomb lattice of silicene is flexible enabling band gap engineering in this material. Additionally theoretical studies have suggested the presence of massless Dirac Fermions in silicene as in graphene[5]. Although most of the striking properties of graphene have been predicted to occur in silicene, the major advantage of investigating silicene like sheets is that these can be easily incorporated into the present silicon based microelectronics industry. This is expected to lead to development of exotic silicene based nano devices. The interaction of substrate with such atomically thick materials is known to influence the properties of these materials. Therefore experimental investigation of the properties of free standing sheets is highly desirable. We have synthesized free standing up to few layers of silicene like 2D material using non-linear interaction of ultra-fast and ultra-short laser pulses with (111) silicon surface in aqueous ambience. The method leads to production of partially oxidized large area silicene sheets along with quantum dots[6] as by products. The sheets are then separated and treated using hydrazine to remove oxygen functional groups. Atomic scale structural characterizations reveal the hexagonal symmetry of the lattice. X-ray diffraction studies are in good agreement with the electron diffraction results and comparative X-ray diffraction studies suggest that the synthesized sheets are not chunks of (111) silicon exfoliated from the target.

The samples were synthesized using tunable ultrafast (femtosecond) laser based nonlinear process in aqueous medium. The laser wavelength used was selected near the band edge of bulk silicon. In this process, the absorption of photon energy excites the electron to the conduction band; the energy absorbed is then transferred to the phonons resulting in non-thermal melting[7] producing layered silicene like sheets. The samples produced were characterized using high resolution transmission microscopy (HRTEM), high resolution X-ray diffraction (HRXRD), selected area electron diffraction (SAED), Raman, UV-Vis absorption and X-ray Photon spectroscopy (XPS).

First principles calculations were performed using Vienna Ab-initio Simulations Package (VASP) to calculate the structural properties of silicene in low buckled (LB) configuration. Highly efficient ultrasoft pseudopotentials using the exchange correlation of Perdew-Burke-Ernzerhof were employed. A high energy cutoff basis of 550 eV was chosen for the plane wave basis set. The atoms were relaxed using the conjugate gradient algorithm such that the force on each atom was smaller than 0.0001 eV/A. A gamma centered K-points mesh of 23x23x1 was used. The optimized lattice parameters were found to be a=b=3.86Å and the buckling height was found to be ~0.45Å and is in good agreement with previously reported data[5]. The introduction of spin-orbit coupling shows a gap of ~1.9meV.

HRTEM micrographs along with SAED pattern in figure 1, shows the formation of hexagonal honeycomb lattice. The inter-planar distance obtained from these measurement were found to be d = 0.24 nm and 0.08 nm for two hexagons visible in the SAED pattern. The peak at 36.53° obtained using high resolution X-ray diffraction (HRXRD) in the inset figure 1(b) top left pannel is in good agreement with the d-spacing of 0.24 nm. No similar peaks were observed in the comparative study of the HRXRD pattern with that of standard $SiO_2$ data and that of the substrate. Further more the peak at 89.12° in the measured diffraction pattern agrees well with the calculated X-ray diffraction pattern of low buckled silicene. The HRXRD diffraction of the silicon target used showed a peak (100% intensity) at 28.44° which is excellent agreement with the reported data (JCPDS – 01-075-0590) and corresponds to the (111) plane of cubic silicon with Fd-3m space group. Thus the X-ray signature suggests that ultra-fast interaction of ultra-short laser pulses does not simply exfoliate the (111) face of silicon. This however confirms that this non-linear interaction does not simply exfoliate the layers which otherwise may be inferred from synthesis of graphene sheets and quantum dots from graphite targets[8] as graphene does not buckle and is stable in a planar geometry. The processes involved causes a phase change to form 2D hexagonal lattice of silicene and is also observed in synthesis of stanene[9]. Further detailed investigations related to change of phase in such non-linear processes is not within the scope of this manuscript.

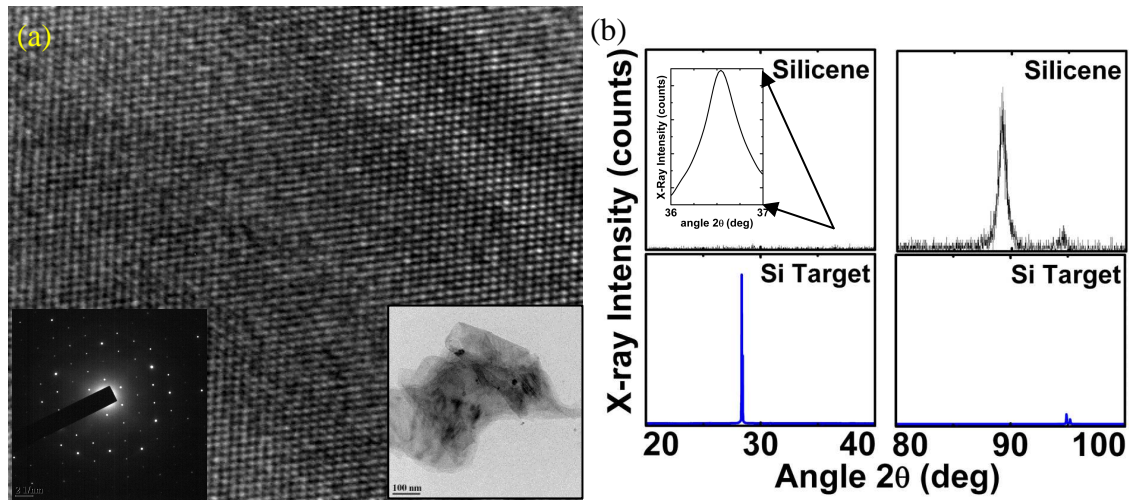

Figure 1:- *(a) HRTEM micrograph of silicene, The left inset below shows the SAED pattern obtained from the sheet suggesting hexagonal symmetry of lattice. The inset on the right below shows a large folded sheet of silicene. (b) Top pannels show HRXRD spectrum of the free standing silicene samples prepared using ultra-fast interaction of ultra-short laser pulses. The inset shows 2θ scan from 36° - 37° in a separate measurement performed on same sample. The bottom pannels show the HRXRD spectrum obtained from the target used.*

The sample was drop-casted on a copper substrate and dried for XPS measurements. The XPS (SPECS surface nano analysis GmbH) measurements were performed using Al-Kα X-ray source. Since silicene is known to oxidize in atmosphere, the exposure leads to oxidation of the sample resulting in formation of oxidized silicene sheets. The XPS spectrum of dried sample on copper substrate in figure 2 shows formation of Si-O bonds due to presence of peak at binding energy ~101.9 eV, moreover traces of backbone of silicene can be observed ~98.4 eV. This is slightly lower as compared to that of elemental silicon and good agreement with the recently reported XPS investigations of silicene sheets grown using MBE growth[10].

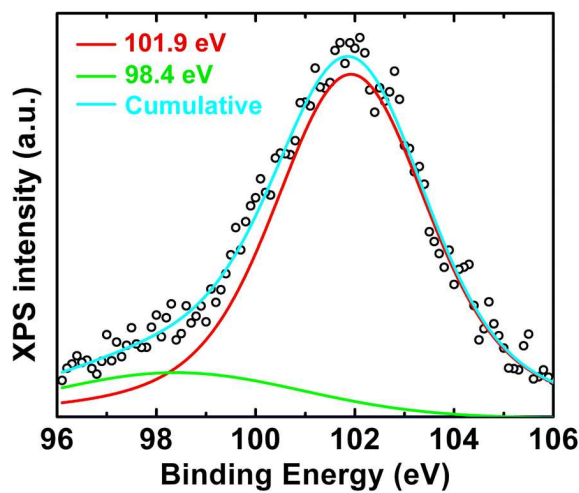

Figure 2:- *XPS spectrum of silicene samples shows cumulative peak at BE of ~101.9 eV and deconvoluted peaks at ~98.4eV and ~101.9eV*

The synthesized samples were further characterized for optical signatures using Raman (Witec alpha 300RAS) and UV-Vis absorption spectroscopy (Shimadzu 2600) and the samples were further checked for the possibility of simple exfoliation of (111) silicon.

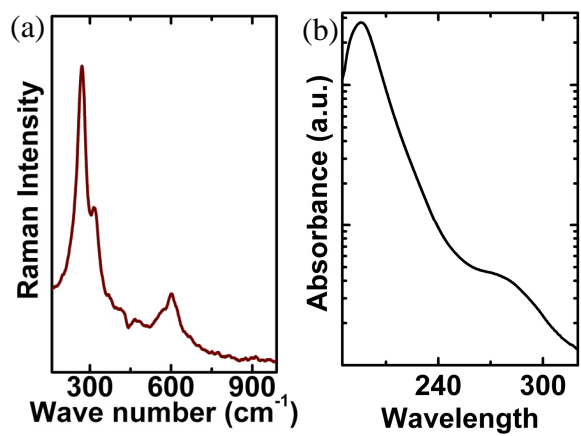

Figure 2:- *(a) µRaman spectrum of silicene samples using excitation wavelength of 532 nm. (b) UV-Vis absorption spectrum of silicene showing absorption peak at ~196nm.*

Bulk silicon gives a sharp 1st order Raman peak at ~ 520 nm using 488nm wavelength of light[11] and ~523 nm using 632.8 nm[12]. Raman spectrum for $2\sqrt{3}\times2\sqrt{3}$ epitaxial superstructures of silicene is observed ~ 521 cm$^{-1}$. The oxidation of these layered superstructures did not show any Raman signature[13]. The Raman spectrum of the synthesized sample shows broad band ~570cm$^{-1}$ – 595cm$^{-1}$. The observed Raman spectrum is in good agreement to the in-plane transversal Optical (iTO) and in-plane longitudinal Optical (iLO) phonon branches predicted using DFT. The origin is similar to the 'G' peak at 1600 cm$^{-1}$ in the Raman spectrum of graphene. Thus this gives a clear indication that the sample synthesized is not an exfoliated (111) chunk of bulk silicon, but likely to be layered silicene like 2D material. The sample also shows defect band ~271cm$^{-1}$ – 315cm$^{-1}$. This is understood to arise from the edges of the sheets as observed in DFT calculations of silicene nanoribbons[14]. The UV-Vis absorption spectrum suggests that silicene shows strong absorption peak at ~196 nm and broadened peak at ~275nm in the UV region of the electromagnetic spectrum. These are also in good agreement with the peak position predicted using density functional theory[15].

      To conclude, we have used non-linear processes involved in ultrafast laser material interaction to synthesize silicene like sheets. The hexagonal symmetry of the lattice is verified using HRTEM and SAED measurements. Comparative study of X-ray diffraction pattern of samples synthesized with target used suggest that samples are not exfoliated (111) chunks of silicene. XPS measurements indicate the presence of silicene backbone in the synthesized sample and Comparison of Raman and UV-Vis absorption spectroscopy with DFT results suggest that synthesized samples are free standing layers of silicene.

**Acknowledgements**

We acknowledge the support of DST-SERI India, grant via sanction order no. DST/TM/SERI/2k10/12/(G) for this research. We are highly thankful to Dr. A. K. Tripathi, Bhabha Atomic research Center, India (fuel cell materials and catalysis section, chemistry division) and for the XPS measurements.